\documentclass[lettersize,journal]{IEEEtran}
\usepackage{amsmath,amsfonts}
\usepackage{algorithmic}
\usepackage{algorithm}
\usepackage{array}
\usepackage[caption=false,font=footnotesize,labelfont=rm,textfont=rm,subrefformat=parens]{subfig}
\usepackage{textcomp}
\usepackage{stfloats}
\usepackage{url}
\usepackage{verbatim}
\usepackage{graphicx}
\usepackage{cite}
\usepackage{multirow}
\usepackage{diagbox}
\usepackage{booktabs}
\usepackage{makecell}

\usepackage[colorlinks  = true,
            linkcolor   = black,
            urlcolor    = magenta,
            anchorcolor = blue,
            bookmarks   = false
            ]{hyperref}

\begin{document}

\title{Quantization Adaptor for Bit-Level Deep Learning-Based Massive MIMO CSI Feedback}

\author{Xudong Zhang, Zhilin Lu, Rui Zeng and Jintao Wang,~\IEEEmembership{Senior Member,~IEEE}
\thanks{
This work was supported in part by Tsinghua University-China Mobile Research Institute Joint Innovation Center. \textit{(Corresponding author: Jintao Wang.)}

The authors are with the Department of Electronic Engineering, Tsinghua University, Beijing 100084, China and Beijing National Research Center for Information Science and Technology (BNRist). (e-mail: zxd22@mails.tsinghua.edu.cn; luzl18@mails.tsinghua.edu.cn; zeng$\_$r17@163.com; wangjintao@tsinghua.edu.cn).
}
}



\maketitle
\begin{abstract}
In massive multiple-input multiple-output (MIMO) systems, the user equipment (UE) needs to feed the channel state information (CSI) back to the base station (BS) for the following beamforming. But the large scale of antennas in massive MIMO systems causes huge feedback overhead. Deep learning (DL) based methods can compress the CSI at the UE and recover it at the BS, which reduces the feedback cost significantly. But the compressed CSI must be quantized into bit streams for transmission. 
In this paper, we propose an adaptor-assisted quantization strategy for bit-level DL-based CSI feedback. 
First, we design a network-aided adaptor and an advanced training scheme to adaptively improve the quantization and reconstruction accuracy.
Moreover, for easy practical employment, we introduce the expert knowledge of data distribution and propose a pluggable and cost-free adaptor scheme. 
Experiments show that compared with the state-of-the-art feedback quantization method, this adaptor-aided quantization strategy can achieve better quantization accuracy and reconstruction performance with less or no additional cost.
The open-source codes are available at \href{https://github.com/zhang-xd18/QCRNet}{https://github.com/zhang-xd18/QCRNet}.
\end{abstract}


\section{Introduction}

\IEEEPARstart{M}{assive} multiple-input multiple-output (MIMO) has been regarded as a promising technology to improve the spectrum and energy efficiency for 5G and beyond systems\cite{larsson2014massive}. 
In massive MIMO systems, the base station (BS) needs to obtain the downlink channel state information (CSI) for beamforming to exploit the potential benefits of MIMO.
However, in frequency division duplexing (FDD) systems, the downlink CSI has to be estimated at the user equipment (UE) and then fed back to the BS due to the lack of channel reciprocity \cite{sim2016compressed}. 
The feedback overhead is unacceptable in massive MIMO systems due to the large scale of antennas. Thus, CSI compression before feedback is especially necessary. 

The traditional methods based on compressed sensing (CS) \cite{kuo2012compressive} require the CSI matrices to be sparse enough. 
However, the practical system may not fully satisfy this requirement especially when the compression ratio is large. 
With the development of artificial intelligence (AI) and deep learning (DL), researchers have been inspired to use neural networks to solve this problem. 
CsiNet \cite{wen2018deep} is the first to introduce the DL-based auto-encoder \cite{vincent2008extracting} architecture into CSI feedback and shows its overwhelming superiority against traditional CS methods. 

The DL-based auto-encoder mainly consists of an encoder and a decoder network.
The CSI matrices estimated at the UE are first pre-processed for feedback. 
Then the encoder network extracts the features and compresses the CSI into a low-dimensional codeword. 
The codeword is transmitted back to the BS through the digital communication channel.
At the BS, the received codeword is recovered to the CSI matrix by the decoder network. 
After some post-processing such as 2D-DFT and zero filling, the original CSI matrix is reconstructed.

In digital communication systems, signals must be transformed into bit streams for feedback.
After compression by the encoder, the codeword consists of full precise float 32 (FP32) numbers. 
Direct transmission of the float numbers will cause unbearable overhead to the feedback channel. 
Thus, the codeword should be quantized first into low bit width before feedback. 
However, the quantization module will introduce quantization errors to the codeword and influence the reconstruction accuracy of the decoder. 
Specific optimization for quantization should be introduced to improve the system performance. 

Therefore, to reduce the impact of quantization, we propose an adaptor-aided quantization strategy in this paper.
The adaptor adjusts values in the codeword and aims at reducing the information loss brought by quantization. 
With more precise codewords, the reconstruction process at the decoder can be improved.
We design two types of adaptors, namely network-based adaptor and expert knowledge-based adaptor. 
Notably, \cite{chen2019novel} proposes the first network-based adaptor, using an offset network before the decoder to improve the quantization accuracy.
However, the network adaptor in \cite{chen2019novel} is too heavy for the resource-sensitive feedback task.
And its training strategy can be further optimized for better performance. 

More specifically, we first design a lightweight network-aided adaptor strategy which achieves better performance with less computation cost. 
Furthermore, we propose a pluggable and cost-free adaptor based on expert knowledge for more practical deployment. 
The main contributions of this paper are listed below.


\begin{itemize}
    \item We design two lightweight adaptor network structures which can achieve excellent performance under indoor and outdoor scenarios. The networks cause less deployment overhead and provide better results compared with the offset network structure \cite{chen2019novel}.
    \item We propose a cost-free expert knowledge-based adaptor considering the practical deployment. 
    We analyze the characteristics of data distribution and use them as expert knowledge to design an adaptor to adjust the data distribution in the codeword.
    The adaptor brings no extra cost and modification to the original network and shows great performance and robustness under different channels.
    \item A novel advanced training scheme is designed to improve the learning ability of the adaptor networks. The training stages and schedulers are meticulously adjusted to fit the network to the task. With the advanced training strategy, the quantization adapting is conducted more sufficiently and the whole feedback performance is improved.
\end{itemize}

The rest of this paper is arranged as follows. Section II lists the related works and section III introduces the system model for bit-level DL-based CSI feedback. Section IV describes the details of the adaptor design, including the network-aided adaptor structure and the pluggable adaptor design. Section V presents the advanced training scheme for adaptor networks. The numerical results and analysis are given in section VI and the conclusion is drawn in section VII. 


\section{Related work}

\begin{figure*}[!t]
    \centering
    \includegraphics[scale=0.88]{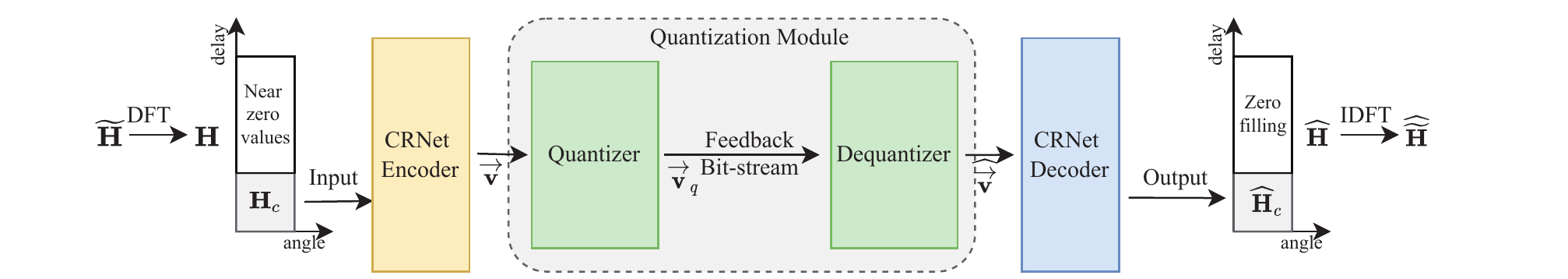}
    \caption{Bit-level DL-based CSI feedback network structure.}
    \label{structure}
\end{figure*}


After CsiNet in \cite{wen2018deep} introduces deep learning into the CSI feedback task, later researchers make much effort to improve the performance of this scheme with advanced network structures and training mechanisms.
CRNet \cite{lu2020multiresolution} proposes a cosine learning rate scheduler which is able to effectively boost training. 
The multi-resolution mechanism is explored in many works \cite{lu2020multiresolution, hu2021mrfnet, xu2021dfecsinet, cheng2021multi} with parallel feature extraction paths. 
Based on the convolutional network, CsiNet+ proposed in \cite{guo2020convolutional} optimizes the convolutional kernels and increases its receptive field. 
CLNet \cite{ji2021clnet} designs a forged complex-valued input layer and increases the feedback performance.
References \cite{cui2022transnet, xu2022transformer} utilize the famous transformer architecture \cite{ashish2017attention} and obtain performance improvement. 
Moreover, as a widely used method in deep learning, the attention mechanism is also introduced to increase the ability of feature extraction in \cite{cai2019attention, song2021saldr}. 
In this way, different parts of CSI matrices can be paid different attention to, and features of higher importance tend to be retained and transmitted. 


In addition to performance improvement, the practical employment of the DL-based feedback scheme is promoted by many researchers. 
Deployment cost is an important factor in practical applications and many works try to reduce it. 
The elastic feedback scheme is proposed in ACRNet \cite{lu2022binarized} to flexibly adjust the network scale according to different resource limitations. 
ShuffleCsiNet \cite{cao2021lightweight} replaces the vanilla convolutional layers with shuffle layers in \cite{zhang2018shufflenet} and reduces the parameters.
Network quantization is also explored in references \cite{guo2020compression,lu2022binarized,lu2021binary} to reduce the parameter storage cost.
In the meanwhile, codeword quantization is also critical to deploying the DL-based system with acceptable transmission overhead.
Uniform quantization is utilized in \cite{lu2019bitlevel, jang2019deep,chang2021deep} to quantize the float numbers in codewords into bit streams.
\cite{fan2021fully} introduces the A-law quantizer and tests the influence of different values of hyper-parameter A on the quantization performance.
In \cite{chen2019novel}, researchers study the codeword data distribution and prove the effectiveness of the $\mu$-law quantizer. 
It also designs an offset network to compensate for the quantization errors.
\cite{zhang2021deep} equips the $\mu$-law quantizer into its system and improves the feedback performance through end-to-end training.
Focusing on the non-differentiable problem brought by the quantization module, \cite{liang2022changeable} designs a differentiable function to approximate the gradients. 




There are many other quantization methods introduced to improve the bit-level system performance.
\cite{ravula2021deep} combines the quantization and channel coding into the design and proposes an entropy coding-based quantization scheme.
In this way, better system performance can be obtained with fewer feedback bits.
A clustering algorithm-based quantization method is proposed in \cite{shen2022Clustering}. 
It utilizes the k-means method with vector quantization to divide codewords into different quantization intervals.
Furthermore, \cite{rizzello2022learning} proposes a nested dropout layer to define an importance ordering and distribute different bit widths to different codewords according to their importance. 
However, the quantization strategies based on vector quantization need to store the codebook at both the UE and BS sides, which brings extra cost.


In summary, codeword quantization is a critical problem considered in promoting the DL-based feedback system into practical deployment.
Current quantization strategies based on quantizer design or the offset network still need to be further optimized to reduce the additional cost and improve performance.
Therefore, this paper proposes an adaptor-aided quantization strategy that achieves better performance with less or even no additional cost. 

\section{System model}

\subsection{Massive MIMO-OFDM system}
In this paper, we consider a single-cell massive MIMO-OFDM system with $N_{t} >> 1$ antennas at the BS and a single antenna at the UE. 
Orthogonal frequency division multiplexing (OFDM) is employed in this system with $\Tilde{N}_c$ sub-carriers. The received signal at the $n^{th}$ sub-carrier $y_{n} \in \mathbb{C}$ can be derived as follows:
\begin{equation}
\label{system model}
y_{n} = \mathbf{h}_n^{H} \mathbf{s}_{n} x_{n} + z_{n},
\end{equation}
where $\mathbf{h}_{n} \in \mathbb{C}^{N_{t} \times 1}$, $\mathbf{s}_{n} \in \mathbb{C}^{N_{t} \times 1}$, $x_{n} \in \mathbb{C}$, and $z_{n} \in \mathbb{C}$ denote the channel vector, precoding vector, data-bearing symbol, and additive noise at the $n^{th}$ sub-carrier, respectively. $(\cdot)^H$ represents conjugate transpose.
Let $\Tilde{\mathbf{H}} = \left[\mathbf{h}_{1},...,\mathbf{h}_{\Tilde{N}_{c}}\right]^{H} \in \mathbb{C}^{\Tilde{N}_{c}\times N_{t}}$ be the CSI matrix, which is essential for the BS to design the precoding vectors $\{\mathbf{s}_{1},...,\mathbf{s}_{N_{t}}\}$ effectively. In the FDD system, the UE has to return the downlink CSI $\Tilde{\mathbf{H}}$ to the BS for lack of reciprocity. However, the $2 \Tilde{N}_c N_{t}$ scale of real elements in $\Tilde{\mathbf{H}}$ is unacceptably large for bit-level digital transmission and demonstrates the necessity of CSI compression.

With the channel sparsity in the angular-delay domain, the feedback overhead can be first reduced as follows:
\begin{equation}
\label{FFT}
\mathbf{H} = \mathbf{F}_d \Tilde{\mathbf{H}} \mathbf{F}_{a}^{H},
\end{equation}
where $\mathbf{F}_d$ and $\mathbf{F}_{a}$ are the $\Tilde{N}_c\times\Tilde{N}_c$ and $N_t \times N_t$ DFT matrices, respectively. For the angular-delay domain channel matrix $\mathbf{H}$, only the first $N_c (< \Tilde{N}_c)$ rows contain large values. The values in the rest of the rows are near zero and can be left out without much loss of information. We use $\mathbf{H}_c$ to denote the first $N_c$ rows of $\mathbf{H}$, which only contains $2 N_c N_t$ elements but is still large for feedback. Therefore, DL-based CSI feedback networks are designed to further compress the CSI matrix.

\subsection{DL-based CSI feedback}
In this paper, we consider the bit-level DL-based CSI feedback network structure as shown in Fig. \ref{structure}.
A widely referenced open-source network named CRNet \cite{lu2020multiresolution} is utilized for the encoder and decoder.
The CSI matrix $\Tilde{\mathbf{H}}$ is first transferred into the angular-delay domain through DFT as (\ref{FFT}) and then reduced to $\mathbf{H}_c$ to feed into the encoder.  $\mathbf{H}_c$ is then feature-extracted and compressed into an $M \times 1$ codeword, where each element will be quantized into $B$ bits to be transmitted back to the BS. It can be described as (\ref{encoder}).
\begin{equation}
\label{encoder}
\mathbf{v}_q = \mathcal{Q}(\mathbf{v}) = \mathcal{Q}(f_{en}(\mathbf{H}_c, \Theta_{en})),
\end{equation}
where $f_{en}(\cdot)$ and $\mathcal{Q}(\cdot)$ denote the encoding and quantizing process, and $\mathbf{H}_c$, $\mathbf{v}$, $\mathbf{v}_q$, and $\Theta_{en}$ are the CSI matrix, codeword, quantized codeword, and parameters of the encoder network, respectively. The compression ratio (CR) can be calculated as follows:
\begin{equation}
\label{cr}
\text{CR} = \frac{2 N_c N_t}{M}.
\end{equation}

Once the BS receives the quantized codeword $\mathbf{v}_q$, the dequantization module and decoder will reconstruct the channel matrix $\hat{\mathbf{H}}$. The recovered channel matrix can be described as follows:
\begin{equation}
\label{decoder}
\hat{\mathbf{H}}_c = f_{de}(\mathcal{D}(\mathbf{v}_q), \Theta_{de}),
\end{equation}
where $f_{de}({\cdot})$ and $\mathcal{D}({\cdot})$ represent the decoding and dequantization process. $\Theta_{de}$ represents the parameters of the decoder network.

\subsection{Quantization module}
In this paper, we utilize the $\mu$-law quantizer which proved to perform well in CSI feedback tasks \cite{chen2019novel}. The $\mu$-law quantizer is a kind of non-uniform quantization. Through a non-decreasing companding function $\Phi:(0,1) \to (0,1)$, it provides narrower quantization intervals near zero and wider intervals near one, thus performing well in operating zero-centralized data. Experiments show that the compressed codewords in CSI feedback tasks are concentrated near zero. The companding function in $\mu$-law can be described as follows:
\begin{equation}
\label{u-law}
\Phi(x) = \frac{\ln(1+\mu |x|)}{\ln(1+\mu)},
\end{equation}
where $x$ is the normalized input and $\mu$ is a parameter to change the degree of companding and should be adjusted according to the distribution of given data.

When operating the $\mu$-law quantizer, the normalized data is first put into the companding function $\Phi$ and quantized by a uniform quantizer, then expanded through function $\Phi^{-1}$ to get the final quantized value. In practical deployment, the companding function $\Phi$ can be approximated by the segmented polyline function, which greatly reduces the computing complexity. 
In this work, we use this $\mu$-law quantizer as a basic quantization method and further propose an adaptor-aided quantization strategy to adjust the codewords for better quantization and reconstruction performance.

\section{CSI feedback with quantization adaptor}

\subsection{Bottleneck FC adaptor design}
In this section, we propose a bottleneck FC adaptor network to adapt the quantized codewords and reduce the quantization errors.
The fully-connected layers are widely used as the feature-extracting module. 
A residual connection structure in \cite{he2016deep} is used to make the adaptor learn and bridge the gap between original codewords and quantized codewords. In the meanwhile, the residual structure can also avoid overfitting. 

\begin{figure}[htb]
    \centering
    \includegraphics[width=\linewidth]{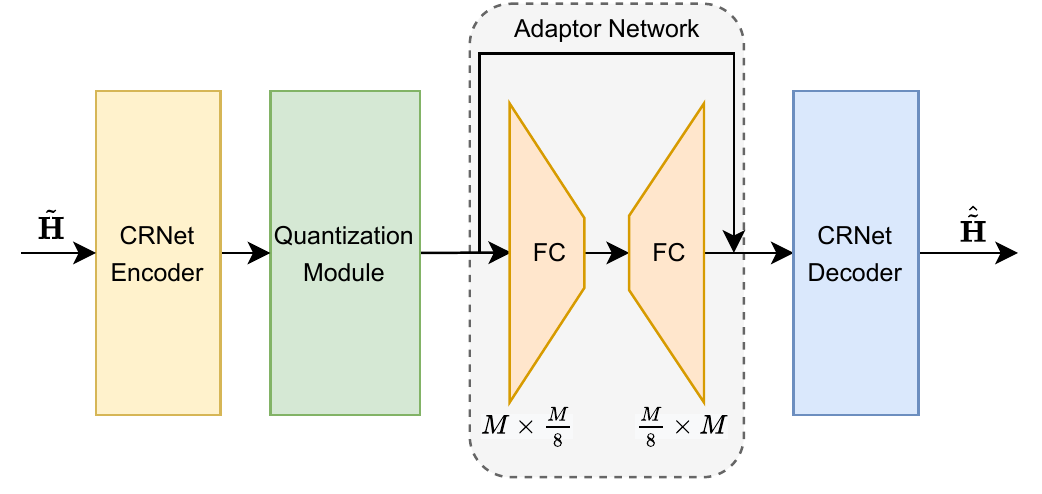}
    \caption{Structure of the bottleneck FC adaptor.}
    \label{bottleneck}
\end{figure}

Considering the practical employment, the FC layers occupy most of the parameter overhead. Thus we utilize the bottleneck structure, in which the middle channels are fewer than the input and output ones. This structure can alleviate the heavy overhead brought by FC layers. Specifically, a 1/8 bottleneck is used for the bottleneck adaptor, and the parameter cost is reduced from $3M^2$ to $M^2/4$ compared with the OffsetNet benchmark in \cite{chen2019novel}. The detailed structure is shown in Fig.\ref{bottleneck}.

In the feedback process, the received quantized codeword is first dequantized and then post-processed by the adaptor network. 
The characteristics of quantization noise can be learned by the adaptor through sufficient training so that the errors in codewords are compensated. 
There is less information loss in the codewords passed into the decoder and the reconstruction performance is then improved.

\subsection{Parallel bottleneck FC adaptor design}

\begin{figure}[tb]
    \centering
    \includegraphics[width=\linewidth]{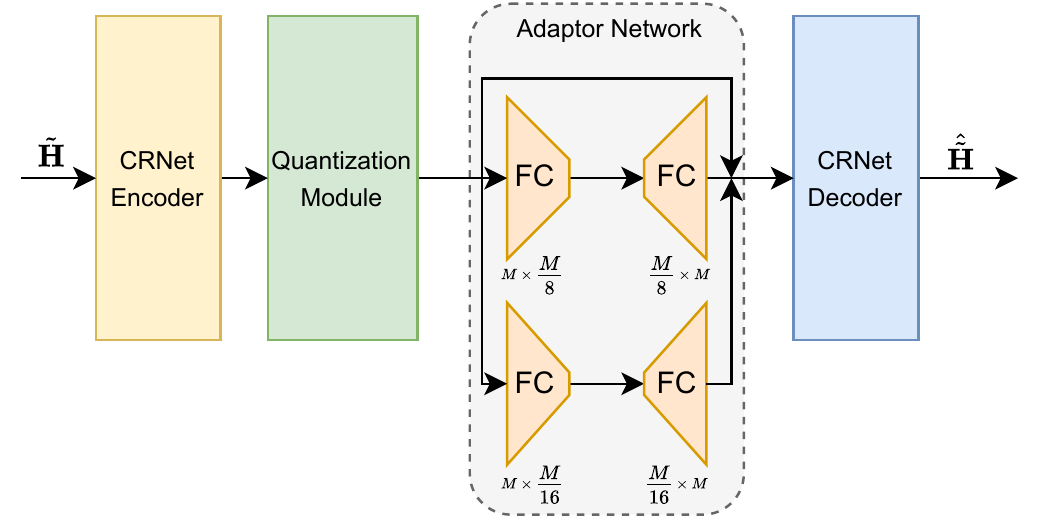}
    \caption{Structure of the parallel bottleneck FC adaptor.}
    \label{parabottleneck}
\end{figure}

In outdoor scenarios, the channel features have more diversity of scales and the matrices can be denser. 
Thus, the parallel branch structure is more suitable to capture the features.
Besides, inspired by \cite{lu2020multiresolution, hu2021mrfnet} which introduce multi-resolutions skills into CSI feedback with different convolutional kernel sizes, we design a parallel bottleneck FC adaptor.
In this adaptor, we add a parallel branch with a different bottleneck scale to the original structure.
Different sizes of bottlenecks can capture features of different scales and further make the recovery more accurate.
This detailed structure is shown in Fig. \ref{parabottleneck}.

Despite the additional branch, the total parameter cost can still be saved a lot compared with the OffsetNet benchmark \cite{chen2019novel} owing to the bottleneck structure. 
Specifically, with two bottlenecks of $1/8$ and $1/16$ sizes, the total parameters are reduced from $3M^2$ to $3M^2/8$, which reveals the validity of bottleneck structures in reducing parameters.

\subsection{L1 adaptor design}
\begin{figure}[tb]
    \centering
    \includegraphics[width=0.8\linewidth]{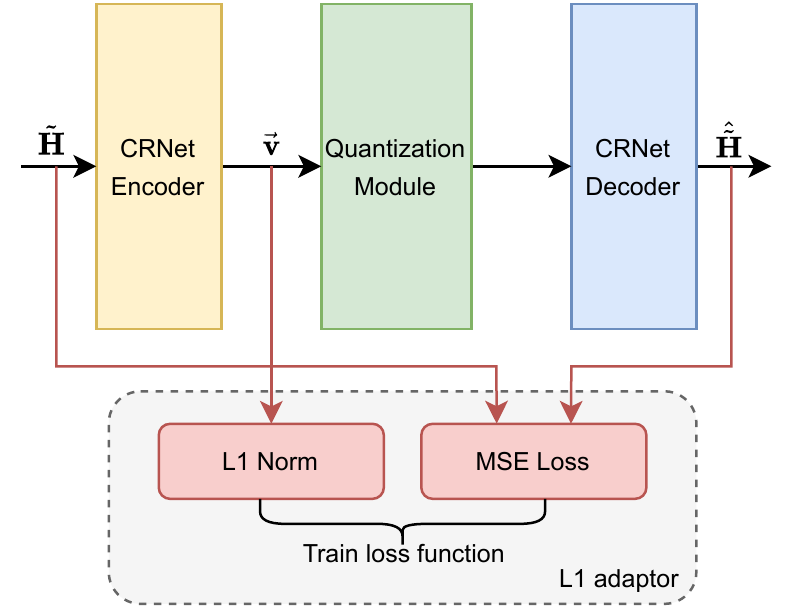}
    \caption{Structure of the L1 adaptor.}
    \label{L1adaptor}
\end{figure}

Although the two bottleneck adaptor structures above can greatly reduce the deployment overhead, extra costs still exist which is not friendly to the resource-sensitive environment.
Therefore, we further propose a totally cost-free adaptor design called L1 adaptor.
This idea mainly comes from the knowledge that quantization performance depends on whether the characteristics of the quantizer match the data distribution.
For the $\mu$-law quantizer, the quantization performance is better when the quantized data distribute close to zero.
Therefore, we design the L1 adaptor to gather the compressed CSI codewords close to zero. 
As shown in Fig. \ref{L1adaptor}, we add an L1 norm to the loss function, which can reduce the norm of the codeword and make the values near zero through training.
In this way, the distribution of codewords is better aligned with the demand of the $\mu$-law quantizer, and better quantization performance is achieved. 
The L1 norm term can be described as $L_1(\mathbf{x}) = ||\mathbf{x}||_1 = \sum_{i=1}^{M} |x_i|$, where $\mathbf{x} \in \mathbb{R}^{M\times1}$ is the normalized data with each element in $[-1,1]$, and $|\cdot|$ represents the absolute value.

With the L1 adaptor, the whole loss function includes two parts: the mean square error (MSE) loss between $\hat{\mathbf{H}}_c$ and $\mathbf{H}_c$ and the L1 norm of the normalized compressed codeword $\mathbf{v}$. The whole loss function can be described as follows:

\begin{equation}
\label{L1 loss}
L\left(\Theta_{en}, \Theta_{de}\right) = \frac{1}{N} \sum_{i=1}^{N} ||{\hat{\mathbf{H}}}_{ci} - \mathbf{H}_{ci} ||_2^2 + || \vec{\mathbf{v}_i} ||_1,
\end{equation}
where $\Theta_{en}$ and $\Theta_{de}$ represent the parameters of the encoder and decoder. $N$ is the number of training samples. $\hat{\mathbf{H}}_{ci}$, $\mathbf{H}_{ci}$, and $\vec{\mathbf{v}_i}$ represent the recovered CSI matrix, original CSI matrix, and compressed codeword, respectively.

Experiment results show that the L1 adaptor can reduce the quantization distortion and further improve the reconstruction performance.
It is also robust to different scenarios and channel models which will be discussed in section VI.
Moreover, this L1 adaptor only changes the training process and brings no adjustments to the network structure.
Thus, it is plug-and-play and can be easily applied to a wide range of feedback networks.
It does not bring any additional parameters to the encoder or decoder and is cost-free for inference. Compared with methods of OffsetNet \cite{chen2019novel} or adaptor networks, it has more advantages under resource-limited conditions.

\section{Training strategy for network adaptor}

\subsection{Two-stage training pipeline design}
In this section, we introduce a two-stage training pipeline to improve the reconstruction performance and the quantization accuracy.

The first stage is to train the whole network parameters with the mean square error (MSE) loss as follows:
\begin{equation}
\label{1-stage loss}
L_{MSE}(\Theta) = \frac{1}{N} \sum_{i=1}^{N} ||\hat{\mathbf{H}}_{ci} - \mathbf{H}_{ci} ||_2^2,
\end{equation}
where $\Theta = \left\{ \Theta_{en}, \Theta_{NA}, \Theta_{de} \right\}$ and $N$ represent the parameters of the whole network and the number of training samples. $\hat{\mathbf{H}}_{ci}$ and $\mathbf{H}_{ci}$ represent the recovered CSI matrix and original CSI matrix. 

With this stage, the whole network is first trained to capture and compress the channel features. 
Through sufficient training, the network will converge to a stage with enough compression and reconstruction ability. 
Further performance optimization can be conducted on such a basis.

The second stage is mainly about training the adaptor network to adjust the data distribution for quantization error compensation.
The encoder parameters are fixed and only the adaptor and decoder parameters are updated in this stage.
In this way, the learned information from the first stage can be retained as much as possible, avoiding degradation of reconstruction performance. 
The quantization signal-to-noise ratio (QSNR) loss is added to the MSE loss as a regularization term against the quantization errors. 
In this stage, the whole loss function can be described as follows:
\begin{equation}
\label{2-stage loss}
    L(\Theta_{NA},\Theta_{de}) =  \frac{1}{N} \sum_{i=1}^{N} ||\hat{\mathbf{H}}_{ci} - \mathbf{H}_{ci} ||_2^2 + \alpha \frac{|| \vec{\mathbf{v}_i} ||_2^2}{|| \vec{\mathbf{v}_i} - \vec{\mathbf{v}_{qi}} ||_2^2},
\end{equation}
where $\Theta_{NA}$ and $\Theta_{de}$ represent the parameters of the adaptor and decoder. $N$ is the number of training samples. $\hat{\mathbf{H}}_{ci}$, $\mathbf{H}_{ci}$, $\vec{\mathbf{v}_i}$, and $\vec{\mathbf{v}_{qi}}$ represent the recovered CSI matrix, original CSI matrix, compressed codeword, and quantized compressed codeword of the $i^{th}$ sample, respectively. The parameter $\alpha$ is the regularization weight to adjust the influence of quantization improvement. Its scheduler design is described in the following sub-section.

After the second stage of training, the network can achieve better feedback performance through the quantization error compensation learned by the adaptor network. 
Experiments show that without the first training stage, the network can not converge to a well-trained state with sufficient compression ability. 

\subsection{Quantization adapting factor scheduler design}

The quantization adapting factor $\alpha$ is a hyper-parameter to adjust the impact of QSNR loss.
An overlarge factor will destroy the reconstruction performance, while an excessively small factor can not fully improve the quantization performance.
In this section, we design a scheduler for $\alpha$, making the quantization adapting factor vary through the training process and work sufficiently in the middle of the training. 

\begin{figure}[htb]
    \centering
    \includegraphics[width=\linewidth]{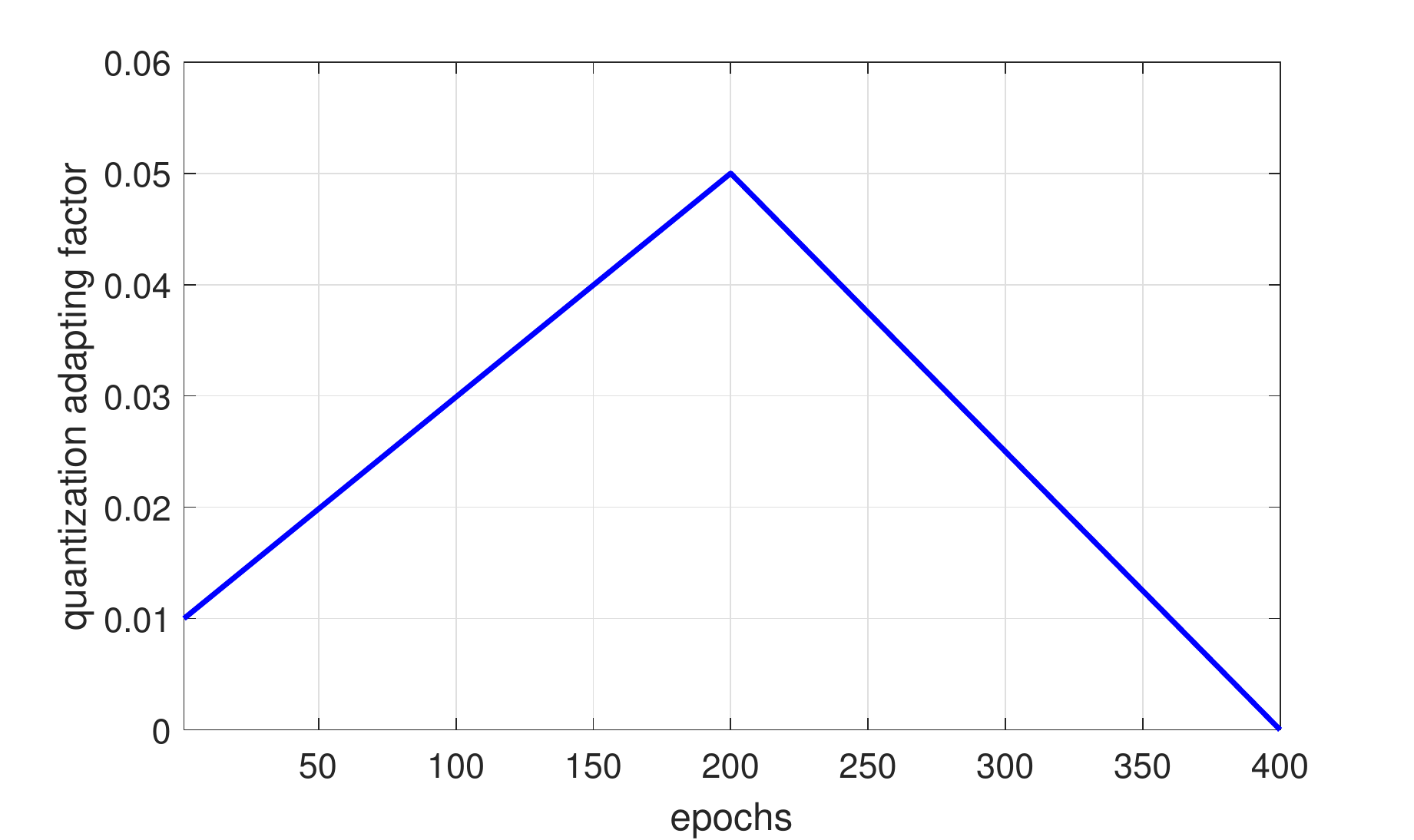}
    \caption{The quantization adapting factor scheduler.}
    \label{ascheduler}
\end{figure}

As shown in Fig. \ref{ascheduler}, $\alpha$ is first linearly increased from 0.01 to 0.05, and then linearly decreased to zero in the end. 
The factor starts from a relatively small value to keep the performance of the original network. 
As the factor is increasing, the QSNR regularization term becomes larger and makes the network converge in the direction of reducing quantization errors. 
After the peak, $\alpha$ starts to decrease and a chance is provided for the network to finetune and get better performance on the basis of the previous adjustment about quantization. 
In this way, the quantization adapting can fully influence the training process and bring benefits to CSI compression and reconstruction. 
Ablation studies about the scheduler design are presented in the next section.

\section{Simulation results and analysis}

\subsection{Simulation settings}

We follow the channel model with default settings in COST2100 \cite{liu2012cost}, which is a widely used dataset in CSI feedback. 
We consider both the indoor scenario at 5.3GHz and the outdoor scenario at 300MHz. 
$N_t = 32$ transmit antennas in uniform linear array (ULA) at the BS are considered. 
We take $\Tilde{N}_c = 1024$ sub-carriers for OFDM and only the first $N_c = 32$ rows of CSI matrix are reserved. 
150,000 independently generated channel samples are split into three parts. The training, validation, and test dataset contain 100,000, 30,000, and 20,000 matrices, respectively. The batch size is set to 200.

Experiments are implemented using the Pytorch library. 
We adopt the same encoder and decoder structure and cosine annealing scheduler for the learning rate proposed in CRNet \cite{lu2020multiresolution}.
For the network-aided adaptor scheme, the network is trained for 1000 epochs for the first stage and 400 epochs for the second stage. 
For the L1 adaptor scheme, the network is only trained for 1000 epochs.

We choose the pure $\mu$-law quantizer and OffsetNet \cite{chen2019novel} as the baselines.
OffsetNet in \cite{chen2019novel} is a three-layer FC module with a residual connection.
For a fair comparison, we deploy the same offset network into the CRNet \cite{lu2020multiresolution} autoencoder. All other settings are the same as the original paper \cite{chen2019novel}. 
We consider the CR of 4, 8, and 16 and the quantization bit number B of 4 and 6. The parameter $\mu$ in the companding function of the $\mu$-law quantizer is set to 50. 

For clearer representation, we provide some notations for different methods here. 
We note the pure $\mu$-law quantizer and the offset network method \cite{chen2019novel} as $\mu$-law and OffsetNet.
Our proposed bottleneck FC adaptor, parallel bottleneck FC adaptor, and L1 adaptor are denoted as BottleFC, ParaBottleFC, and L1Adaptor, respectively.



\subsection{Quantization performance}

We propose the quantization performance of different methods in this part.
For evaluation, we compute the quantization signal-to-noise ratio (QSNR) as the criterion,
\begin{equation}
\label{QSNR}
\text{QSNR} = \mathbb{E}\left[\frac{||\mathbf{v}||_2^2}{||\mathbf{v}-\mathbf{v}_q||_2^2}\right],
\end{equation}
where $\mathbf{v}$ and $\mathbf{v}_q$ denote the compressed CSI codeword before and after quantization. $|| \cdot ||_2$ denotes the Euclidean norm.

\begin{table}[b]
\caption{SNR (dB) Performance of Different Quantization Strategies in Different Scenarios.\label{SNRs table}}
\centering
\resizebox{\linewidth}{!}{
\begin{tabular}{ c|c|c c c|c c c }
\hline
\hline
\multicolumn{2}{c|}{\textbf{B}}  & \multicolumn{3}{c|}{6bit}  & \multicolumn{3}{c}{4bit}  \\
\hline
\textbf{Scenario} & \diagbox{\textbf{Method}}{\textbf{CR}} & 4 & 8 & 16 & 4 & 8 & 16\\
\hline
\multirow{4}{*}{in} 
 & $\mu$-law & 44.15 & 44.27 & 43.29 & 30.01 & 29.50 & 29.14\\
 & OffsetNet\cite{chen2019novel} & 45.05 & 44.68 & 43.09 & 30.58 & 29.76 & 29.18\\
 & L1Adaptor & \textbf{47.14} & 46.06 & \textbf{44.42} & 30.58 & \textbf{33.14} & \textbf{31.66} \\
 & BottleFC & 47.11 & \textbf{46.19} & 43.62 & \textbf{30.76} & 29.71 & 29.02 \\
\hline
\multirow{4}{*}{out} 
 & $\mu$-law & 35.25 & 36.02 & 37.64 & 23.81 & 22.62 & 24.77 \\
 & OffsetNet\cite{chen2019novel} & 35.74 & 36.84 & 37.74 & 24.27 & 23.08 & 25.07\\
 & L1Adaptor & \textbf{49.28} & \textbf{44.70} & \textbf{41.56} & \textbf{28.46} & \textbf{29.90} & 26.11 \\
 & ParaBottleFC & 39.06 & 36.73 & 35.50 & 27.16 & 27.00 & \textbf{26.65}\\
\hline
\hline
\end{tabular}}
\end{table}

Table \ref{SNRs table} lists the QSNR for different methods under different scenarios. 
We can conclude that the adaptor scheme achieves better performance than baselines in most experimental scenarios. Through data adapting, the quantization errors can be compensated, leading to higher QSNR. 
Moreover, the L1 adaptor achieves better quantization performance than network adaptors in most scenarios. 
The main reason is that the L1 adaptor strictly constrains the data distribution ahead of the quantizer. 
Such early-stage data restriction is more beneficial to the QSNR performance compared with the network-based error compensation after the vanilla quantization process.

\begin{figure*}[!t]
\centering
\includegraphics[width=\linewidth]{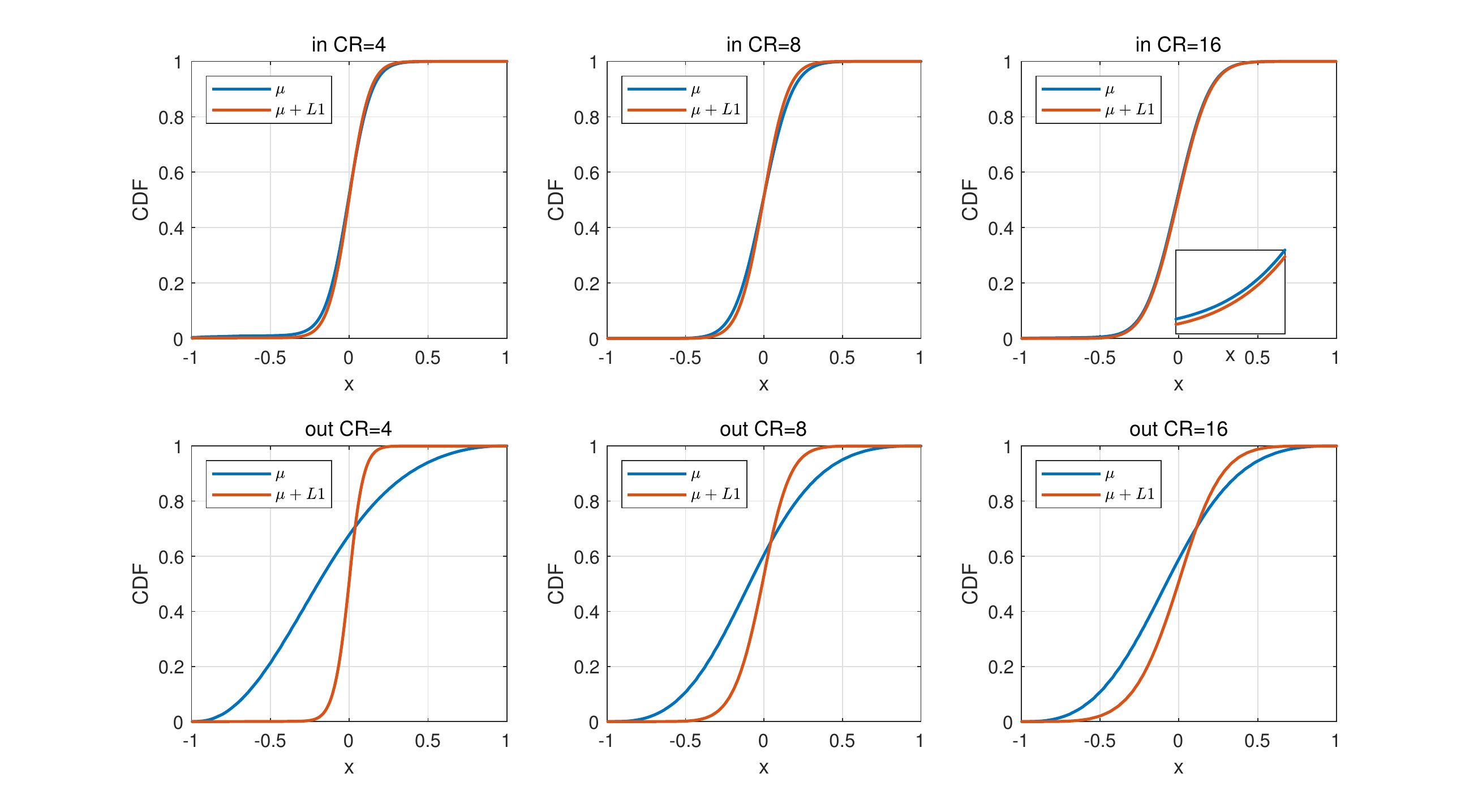}
\caption{Cumulative distribution functions of the compressed CSI codewords in different scenarios with the quantization bit width set to 6. The data are distributed in [-1,1] and centralized near zero. With the L1 adaptor, the cumulative distribution function of data increases more steeply at the near-zero zone.}
\label{data distribution fig}
\end{figure*}

For further clarification, we present the cumulative distribution functions (CDFs) of compressed CSI codewords with or without the L1 adaptor.
The quantization bit width is set to 6 and functions under different scenarios are provided as shown in Fig. \ref{data distribution fig}.
The x-axis denotes the compressed CSI value and the y-axis denotes the cumulative distribution probability of the corresponding value.
The CDFs mainly increase at zero which proves the zero-centralized distribution of CSI codewords.
Compared with outdoor scenarios, CDFs in indoor scenarios are steeper near zero which means CSI codewords in indoor scenarios distribute more concentratedly.
Drawn from the figure, we can conclude that the codewords distribute more close to zero with the L1 adaptor.
This further makes the data distribution and the $\mu$-law quantizer more compatible, thereby bringing quantization performance gain.

\subsection{Reconstruction performance}

The whole system is evaluated with the reconstruction performance of CSI feedback.
We consider the distance between the original CSI matrix ${\mathbf{H}}_c$ and the reconstructed $\hat{\mathbf{H}}_c$ with normalized mean square error (NMSE) defined as follows:

\begin{equation}
\label{NMSE}
\text{NMSE} = \mathbb{E}\left[\frac{||\mathbf{H}_c - \hat{\mathbf{H}}_c||_2^2}{||\mathbf{H}_c||^2_2}\right].
\end{equation}

The main NMSE performance comparison between different methods is given in Table \ref{NMSE performance}. 
Note that NQ represents the feedback with no quantization module. 

\begin{table}[tb]
\caption{NMSE (dB) Performance of Different Quantization Strategies in Different Scenarios.\label{NMSE performance}}
\centering
\resizebox{\linewidth}{!}{
\begin{tabular}{ c|c| c c | c c | c c }
\hline
\hline
\multicolumn{2}{ c|}{\textbf{CR}}  & \multicolumn{2}{ c|}{4}  & \multicolumn{2}{ c|}{8} & \multicolumn{2}{ c }{16}\\
\hline
\textbf{Scenario} & \diagbox{\textbf{Method}}{\textbf{B}} & 6bit & 4bit & 6bit & 4bit & 6bit & 4bit\\
\hline
\multirow{5}{*}{in} & NQ & \multicolumn{2}{c|}{-26.99} & \multicolumn{2}{ c|}{-16.01} & \multicolumn{2}{ c }{-11.35} \\
 & $\mu$-law & -21.92 & -18.45 & -14.39 & -13.70 & -11.17 & -9.768\\
 & OffsetNet\cite{chen2019novel} & -23.23 & -18.61 & -15.08 & -13.82 & -11.25 & -10.31 \\
 & L1Adaptor & -25.25 & -18.71 & -15.38 & -13.95 & -11.61 & -10.48 \\
 & BottleFC & \textbf{-25.82} & \textbf{-19.94} & \textbf{-15.58} & \textbf{-14.32} & \textbf{-11.68} & \textbf{-10.55} \\
\hline
\multirow{5}{*}{out} & NQ & \multicolumn{2}{ c|}{-12.70} & \multicolumn{2}{ c|}{-8.04} & \multicolumn{2}{ c }{-5.44} \\
 & $\mu$-law & -11.51 & -11.31 & -7.884 & -7.873 & -5.259 & -5.139 \\
 & OffsetNet\cite{chen2019novel} & -11.61 & -11.48 & -8.074 & -7.938 & -5.301 & -5.250 \\
 & L1Adaptor & -11.79 & -11.69 & -8.223 & -8.045 & -5.437 & -5.287 \\
 & ParaBottleFC & \textbf{-12.70} & \textbf{-12.13} & \textbf{-8.352} & \textbf{-8.102} & \textbf{-5.513} & \textbf{-5.359} \\
\hline
\hline
\end{tabular}}
\end{table}

From Table \ref{NMSE performance}, the adaptor scheme achieves better performance than baselines in all scenarios. 
The better quantization accuracy brought by adaptors promotes the decoding and recovery process. 
Moreover, according to Table \ref{NMSE performance}, the effect of reconstruction is more sensitive to the quantization bit width and strategy in indoor scenarios than in outdoor scenarios.
This is because the complex features of outdoor channels perform as the dominant challenge of feature extracting and compressing in feedback. 

Besides, the network-aided adaptor scheme outperforms the L1 adaptor although the L1 adaptor achieves higher QSNR performance. 
Compared with hard constraints brought by the L1 adaptor, the learnable network adaptor can balance its attention paid to the quantization improvement and recovery performance. 
Thanks to the two-stage training pipeline and $\alpha$ scheduler, the network adaptor is able to optimize the feedback NMSE better under the pressure of quantization regularization. 
As we can see, the proposed network adaptors achieve the best performance under both indoor and outdoor scenarios. 
It is worth mentioning that in some conditions, its performance even approaches or exceeds the no-quantization performance, which reveals the effectiveness of the additional adaptor network. 

The complexity comparison is listed in Table \ref{complexity}. 
The L1 adaptor outperforms the vanilla $\mu$-law scheme without any extra deployment overhead, making it a practical cost-free upgrade for the quantization-aware feedback system.
In the meanwhile, the two network-based adaptors achieve the best reconstruction performance with lower complexity compared with the previous state-of-the-art OffsetNet \cite{chen2019novel}, which proves the superiority of our network adaptor design.

\begin{table}[b]
\footnotesize
\caption{Complexity comparison of different quantization strategies.\label{complexity}}
\centering
\resizebox{\linewidth}{!}{
\begin{tabular}{c| c |c c c c}
\hline
\hline
\textbf{Complexity} & \textbf{CR} & \textbf{OffsetNet}\cite{chen2019novel} & \textbf{BottleFC} & \textbf{ParaBottleFC} & \textbf{L1Adaptor} \\
\hline
\multirow{3}{*}{FLOPs} & 4 & 6.173M &  5.452M &  5.485M &  \textbf{5.386M} \\
 & 8 & 4.534M &  4.354M &  4.362M &  \textbf{4.338M} \\
 & 16 & 3.863M &  3.817M &  3.820M &  \textbf{3.813M} \\
\hline
\multirow{3}{*}{Params Num} & 4 & 2.891M &  2.169M &  2.202M &  \textbf{2.103M} \\
 & 8 & 1.251M &  1.071M &  1.079M &  \textbf{1.054M} \\
 & 16 & 579.1K &  533.8K &  536.0K &  \textbf{529.6K} \\
\hline
\hline
\end{tabular}}
\end{table}

\subsection{Effectiveness of quantization adapting factor scheduler}
In this part, some ablation studies are presented to prove the effectiveness of the quantization adapting factor scheduler. 
The scheduler controls the variation of the adapting factor and further adjusts the weights of QSNR regularization. 
Ablation studies show that the proposed scheduler which increases at first and then decreases works better because such a quantization adaptor mainly limits the network convergence in the middle of the training. 

With thorough experiments, we discover that the feedback performance tends to be better when $\alpha$ lies in the range from 0 to 0.05.
Based on this observation, we further design three types of schedulers.

We consider three types of schedulers. 
First, a constant adapting factor of 0.03 is introduced as a straightforward benchmark scheme. 
Then, a monotonically decreasing scheduler which ranges from 0.05 to 0 is tested to mainly restrict the network convergence at the early training stage. 
Finally, a piecewise linear scheduler as Fig. \ref{ascheduler} shows is proposed with peak $\alpha=0.05$ reached in the middle of the training. 
Compared with the previous linearly decreasing scheduler, we wish to see if the network can further benefit from freer early-stage training.

\begin{figure}[!htbp]
    \centering
    \subfloat[NMSE performance]{\includegraphics[width=\linewidth]{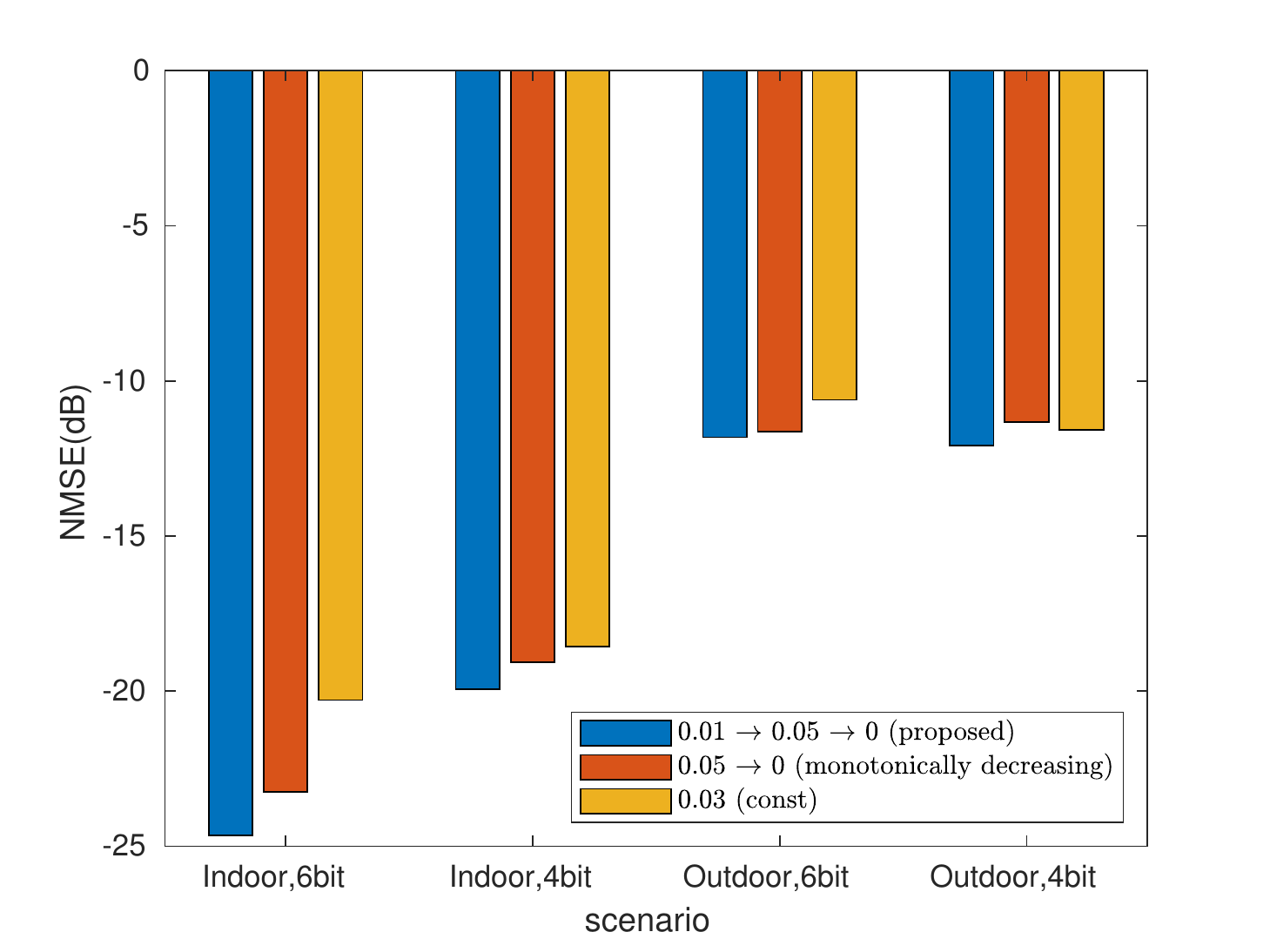}%
    \label{fig:nmse}}
    \hfil
    \subfloat[QSNR performance]{\includegraphics[width=\linewidth]{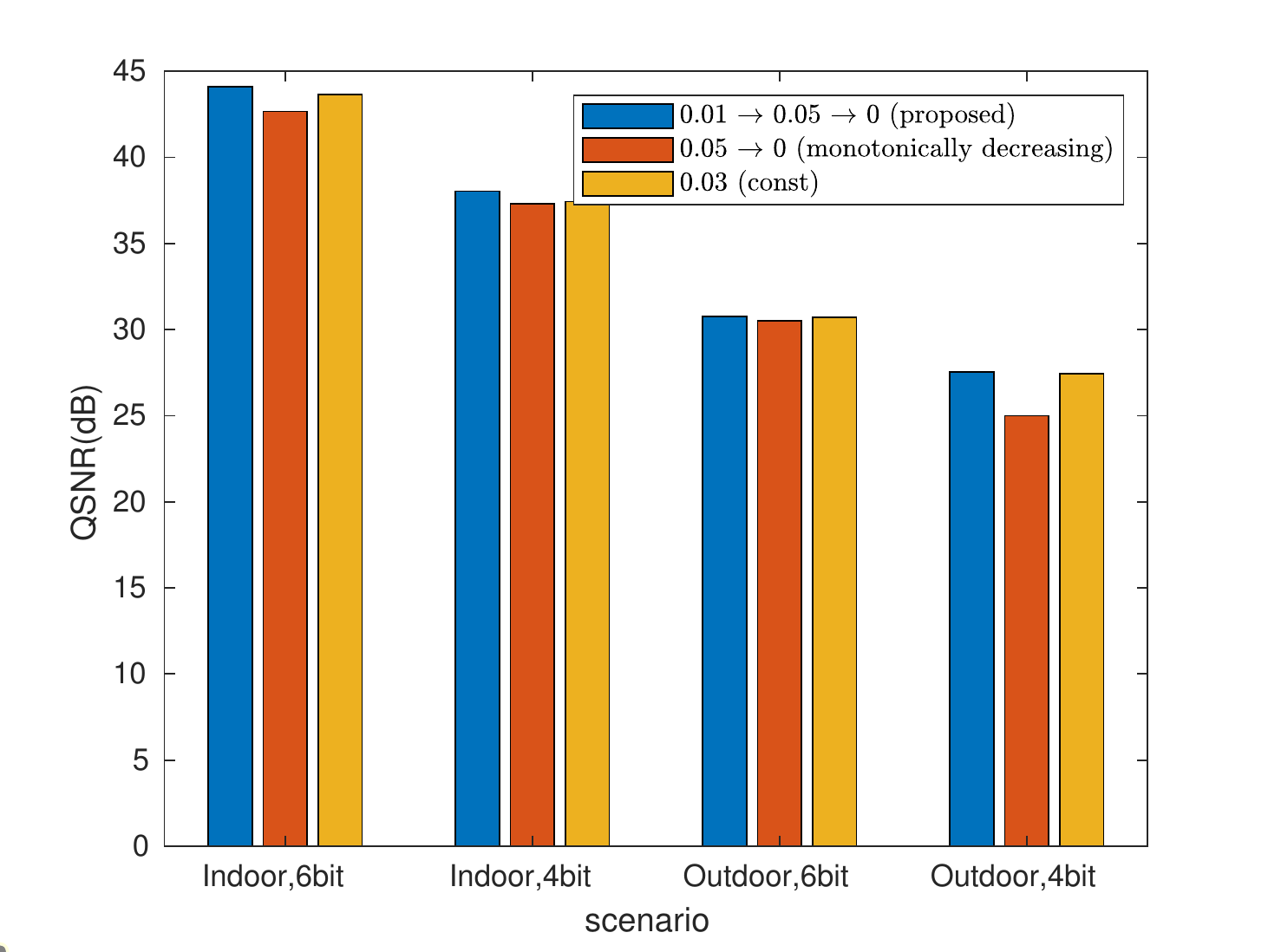}%
    \label{fig:qsnr}}
    \caption{Performance comparison between different quantization adapting schedulers. (a) NMSE performance. (b) QSNR performance.}
    \label{compareofscheduler}
    \end{figure}

Results from Fig. \ref{compareofscheduler} show that the piecewise linear scheduler provides the best quantization and reconstruction performance.
This demonstrates the effectiveness of applying adapting restrictions mainly in the middle stage of training.
It also suggests that too strong constraints in the early stage of training may lead to undermining the abilities learned from the previous stage, thereby reducing the performance of training.

\subsection{Robustness of L1 adaptor}

The L1 adaptor design is derived from the data distribution and may be affected by the communication environment. 
So in this part, we evaluate the performance of the L1 adaptor on another dataset to prove its robustness and effectiveness.
Wireless-Intelligence \cite{wirelessAI} is a public channel dataset library, from which the nrCDLChannel model following 3GPP TR 38.901 \cite{3GPP} standard is used for the CSI generation.
Simulation results show that the L1 adaptor can still obtain solid performance gain when processing the clustered delay line (CDL) channel models.

\subsubsection{Dataset parameters}

The main parameter settings for CDL channel model simulation are listed in Table \ref{CDL}. 

\begin{table}[!b]
    \centering
    \caption{Simulation parameters for CDL channel model.}
    \begin{tabular}{c|c}
    \hline
    \hline
    \textbf{Parameter} & \textbf{Value} \\ 
    \hline
    Carrier frequency & 3.5GHz \\
    Bandwidth & 10MHz \\
    Subcarrier spacing & 15KHz \\
    Subcarriers Num &  1024 \\
    Channel model & CDL-A \\
    Delay spread & 300ns or 30ns \\
    Tx Num $(N_t)$ & 32 \\
    Rx Num $(N_r)$ & 1 \\
    \hline
    \hline
    \end{tabular}
    \label{CDL}
\end{table}

\begin{figure}
    \centering
    \subfloat[]{\includegraphics[width=0.5\linewidth]{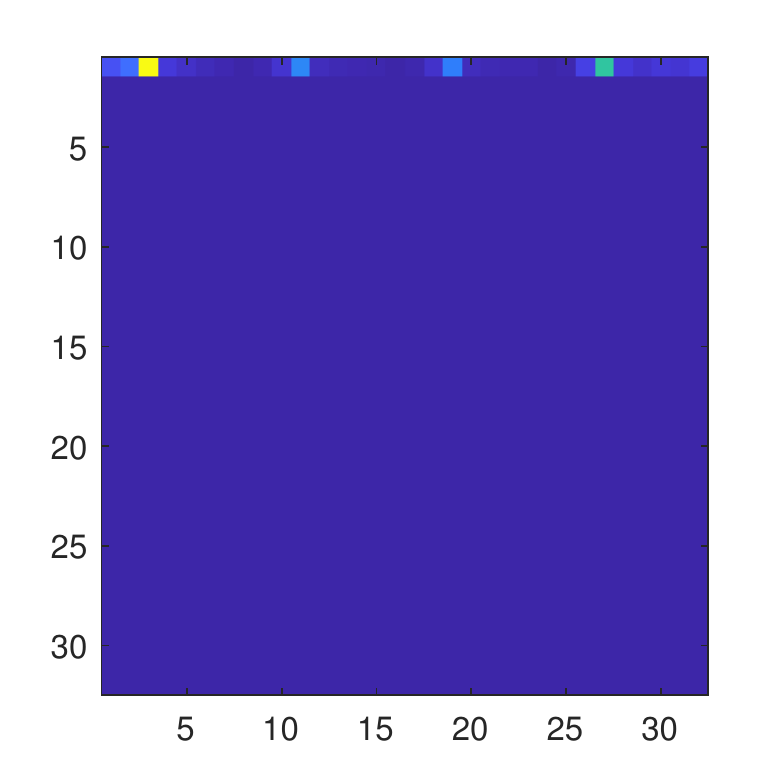}%
    \label{fig:30}}
    \subfloat[]{\includegraphics[width=0.5\linewidth]{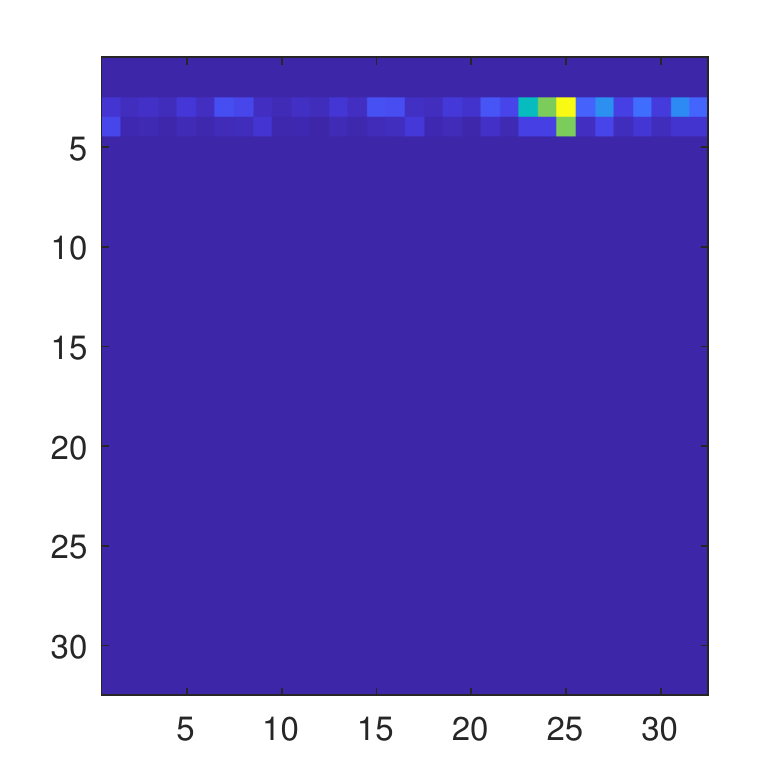}%
    \label{fig:300}}
    \caption{Channel samples of CDL channel models with different delay spread settings. (a) Delay spread = 30ns. (b) Delay spread = 300ns.}
    \label{CDLsample}
    \end{figure}

Fig. \ref{CDLsample} presents two CSI samples in the angular-delay domain of the CDL channel model. 
As shown in the samples, the channel also has sparsity in the angular-delay domain and only the first 32 non-zero rows are meaningful for the feedback task.


\subsubsection{Performance}
In order to examine the robustness of the pluggable L1 adaptor, we test it on CDL channel models. 
Simulation results in Table \ref{CDL performance} show that this adaptor can improve the QSNR and NMSE performance at the same time under different scenarios.
This proves that the data distribution refinement driven by the proposed L1 adaptor takes effect in a robust way. 
In other words, our design is proved to be generally helpful under different channel models and various scenarios.

\begin{table}[tbp]
    \centering
    \caption{QSNR(dB) and NMSE(dB) performance of the L1 adaptor on CDL dataset.}
    \begin{tabular}{c|c|c|c c|c c}
    \hline
    \hline
    \textbf{Delay} & \multirow{2}{*}{\textbf{CR}} & \multirow{2}{*}{\textbf{Methods}} & \multicolumn{2}{c|}{\textbf{NMSE}} & \multicolumn{2}{c}{\textbf{QSNR}$^\mathrm{a}$} \\ 
    \textbf{Spread} & & & 6bit & 4bit & 6bit & 4bit \\
    \hline
    \multirow{9}{*}{30ns} & \multirow{3}{*}{16} & NQ & \multicolumn{2}{c |}{-35.17} & \multicolumn{2}{c}{/} \\
     & & $\mu$-law & -30.44& -21.81& 44.78& 28.87\\
     & & L1Adaptor & \textbf{-30.89}& \textbf{-22.59} & \textbf{45.20}& \textbf{29.26}\\
    \cline{2-7}
     & \multirow{3}{*}{64} & NQ & \multicolumn{2}{c |}{-22.89} & \multicolumn{2}{c}{/} \\
     & & $\mu$-law & -20.63& -16.21& 44.11& 29.30\\
     & & L1Adaptor & \textbf{-21.80}& \textbf{-16.29}& \textbf{44.71} & \textbf{29.65}\\
    \cline{2-7}
     & \multirow{3}{*}{128} & NQ & \multicolumn{2}{c |}{-16.14} & \multicolumn{2}{c}{/} \\
     & & $\mu$-law & -14.68& -12.44& 41.68& 27.04\\
     & & L1Adaptor & \textbf{-15.46}& \textbf{-12.64}& \textbf{44.32}& \textbf{29.72}\\
     \hline
    \multirow{9}{*}{300ns} & \multirow{3}{*}{16} & NQ & \multicolumn{2}{c |}{-31.25} & \multicolumn{2}{c}{/} \\
     & & $\mu$-law & -25.49& -18.94& 45.68& 29.01\\
     & & L1Adaptor & \textbf{-27.68}& \textbf{-18.98} & \textbf{47.50}& \textbf{30.29}\\
     \cline{2-7}
     & \multirow{3}{*}{64} & NQ & \multicolumn{2}{c |}{-15.93} & \multicolumn{2}{c}{/} \\
     & & $\mu$-law & -13.01& -11.50& 42.42& 28.92\\
     & & L1Adaptor & \textbf{-13.86}& \textbf{-12.35}& \textbf{45.99} & \textbf{31.32}\\
    \cline{2-7}
     & \multirow{3}{*}{128} & NQ & \multicolumn{2}{c |}{-11.04} & \multicolumn{2}{c}{/} \\
     & & $\mu$-law & -9.981& -9.482& 40.50& 27.66\\
     & & L1Adaptor & \textbf{-10.33}& \textbf{-9.700}& \textbf{43.53}& \textbf{29.33}\\
    \hline
    \hline
    \multicolumn{7}{l}{$^{\mathrm{a}}  /$ means that the performance is not reported.}
    \end{tabular}
    \label{CDL performance}
    
\end{table}

\section{Conclusion}
In this paper, we proposed an adaptor-aided quantization strategy for DL-based CSI feedback in massive MIMO systems. 
A network adaptor scheme was presented with two lightweight network structures. 
To further reduce the computation cost, we introduced expert knowledge of data distribution and designed a robust L1 adaptor which was totally cost-free and improved the quantization performance through distribution adapting.
Besides, an advanced training pipeline was specially designed to boost the performance.
A quantization adapting factor scheduler was introduced to further optimize the quantization process. 
Experiments show that both adaptor schemes achieved better quantization and reconstruction performance with lower cost and overhead.

\bibliographystyle{ieeetr}
\bibliography{main}

\vfill

\end{document}